\documentclass[twoside,leqno,twocolumn]{article}  
\usepackage{ltexpprt} 
\usepackage{url,graphicx,paralist,amsmath,multirow,subcaption,epsfig}
\begin{document}

\title{Virtual Networks and Poverty Analysis in Senegal}

\author{Neeti Pokhriyal\\
	Computer Science and Engineering\\
	State University of New York at Buffalo\\
	{\tt\small neetipok@buffalo.edu}
	\and
	Wen Dong\\
	Computer Science and Engineering\\
	State University of New York at Buffalo\\
	{\tt\small wendong@buffalo.edu}
	\and
	Venugopal Govindaraju \\
	Computer Science and Engineering\\
	State University of New York at Buffalo\\
	{\tt\small venu@cubs.buffalo.edu}
}
\date{}
\maketitle

\begin{abstract}
Do today's communication technologies hold potential to alleviate poverty? The mobile phone's accessibility and use allows us with an unprecedented volume of data on social interactions, mobility and more. Can this data help us better understand, characterize and alleviate poverty in one of the poorest nations in the world. Our study is an attempt in this direction. We discuss two concepts, which are both interconnected and immensely useful for securing the important link between mobile accessibility and poverty. 

First, we use the cellular-communications data to construct virtual connectivity maps for Senegal, which are then correlated with the poverty indicators to learn a model. Our model predicts poverty index at any spatial resolution. Thus, we generate Poverty Maps for Senegal at an unprecedented finer resolution. Such maps are essential for understanding what characterizes poverty in a certain region, and how it differentiates from other regions, for targeted responses  for the demographic of the population that is most needy. An interesting fact, that is empirically proved by our methodology, is that a large portion of all communication, and economic activity in Senegal is concentrated in Dakar, leaving many other regions marginalized. 

Second, we study how user behavioral statistics, gathered from cellular-communications, correlate with the poverty indicators. Can this relationship be learnt as a model to generate poverty maps at a finer resolution? Surprisingly, this relationship can give us an alternate poverty map, that is solely based on the user behavior. Since poverty is a complex phenomenon, poverty maps showcasing multiple perspectives, such as ours, provide policymakers with better insights for effective responses for poverty eradication. 
\end{abstract}

\section{Introduction and Motivation}

According to the United Nations Development Program's 2014 Human Development Index (HDI), Senegal is ranked 163 out of 187 countries with an HDI index of 0.485. HDI measures achievement in three basic dimensions of human development: health, knowledge, and standard of living. Senegal has a population of 14.1 million, with 43.1\% urban population, and the median age of 18.2 years. It is one of the poorest country in the world, with over 9.2 million people living in multidimensional poverty. Wealth distribution in Senegal is very unequal.

 Poverty incidence remains high, affecting about 47\% of the population. There are wide disparities between poverty in rural areas (at 57\%)s, and urban areas, where the poverty rate is 33\%. More than 42\% of the population lives in rural areas, with a population density that varies from 77 people per square kilometer to 2 people per square kilometer in the dry regions of the country.
 
% A poverty map and its characterization is also an issue of potential impact identified by the D4D committee

On the other hand, the growth in mobile-cellular technology has been very impressive in recent decades. It is estimated that there are 95 mobile-cellular telephone subscriptions per 100 inhabitants worldwide ~\cite{MDG}. In Senegal, there are 93 mobile phone subscriptions per 100 people, according to the latest world-bank report~\cite{cellular_stat}. 

The power of growth of mobile technology poses a question: Can their accessibility be used to identify, characterize, and, in turn, alleviate poverty? Ours is a case study towards answering this question. An expected outcome is a high resolution poverty map of Senegal, and its poverty analysis, with some recommendations for effective policies for an inclusive growth. We believe that such poverty analysis with the growth of virtual mobility will be beneficial to a developing economy like Senegal.
%\begin{figure}[h]
%	\centering
%	\includegraphics[width=0.5\textwidth]{./images/globalICT.jpg}
%	\caption{Global ICT Developments 2000-2014}
%	\label{fig:globalict}
%\end{figure}
\subsection{Poverty Maps}
Currently the poverty maps are created using nationally representative household surveys, which requires a lot of man-power, and time, and continues to lag for Sub-Saharan Africa compared to the world ~\cite{poverty_census_data_availability}. The data is updated yearly, and assessed for poverty progress in 3 years.
%\subsection{Multidimensional Poverty Index}\label{subsec:mpi}

Poverty has traditionally been measured in one dimension, usually income or consumption, called income poverty. Another internationally comparable poverty measure is the World Bank’s \$1.25 per day, which identifies people who do not reach the minimum income poverty line. 

In 2010, the {\em Oxford Poverty and Human Development Initiative} (OPHI) launched a {\em Multidimensional Poverty Index} (MPI). It is a composite of 10 indicators across three areas -- education (years of schooling, school enrollment), health (malnutrition, child mortality), and living conditions. %(including both access to services and proxies for household wealth). 

We use MPI for our poverty analysis, since it closely aligns with the Human Development Index, and is widely accepted to study poverty. MPI is robust to decomposition within relevant sub-groups of populations, like urban vs rural, geographic regions (districts/provinces/states), religion and ethnicity, gender; so that targeted policies can be planned for specific demographics. 

\begin{figure*}[t]
	\centering
	\includegraphics[width=0.7\textwidth]{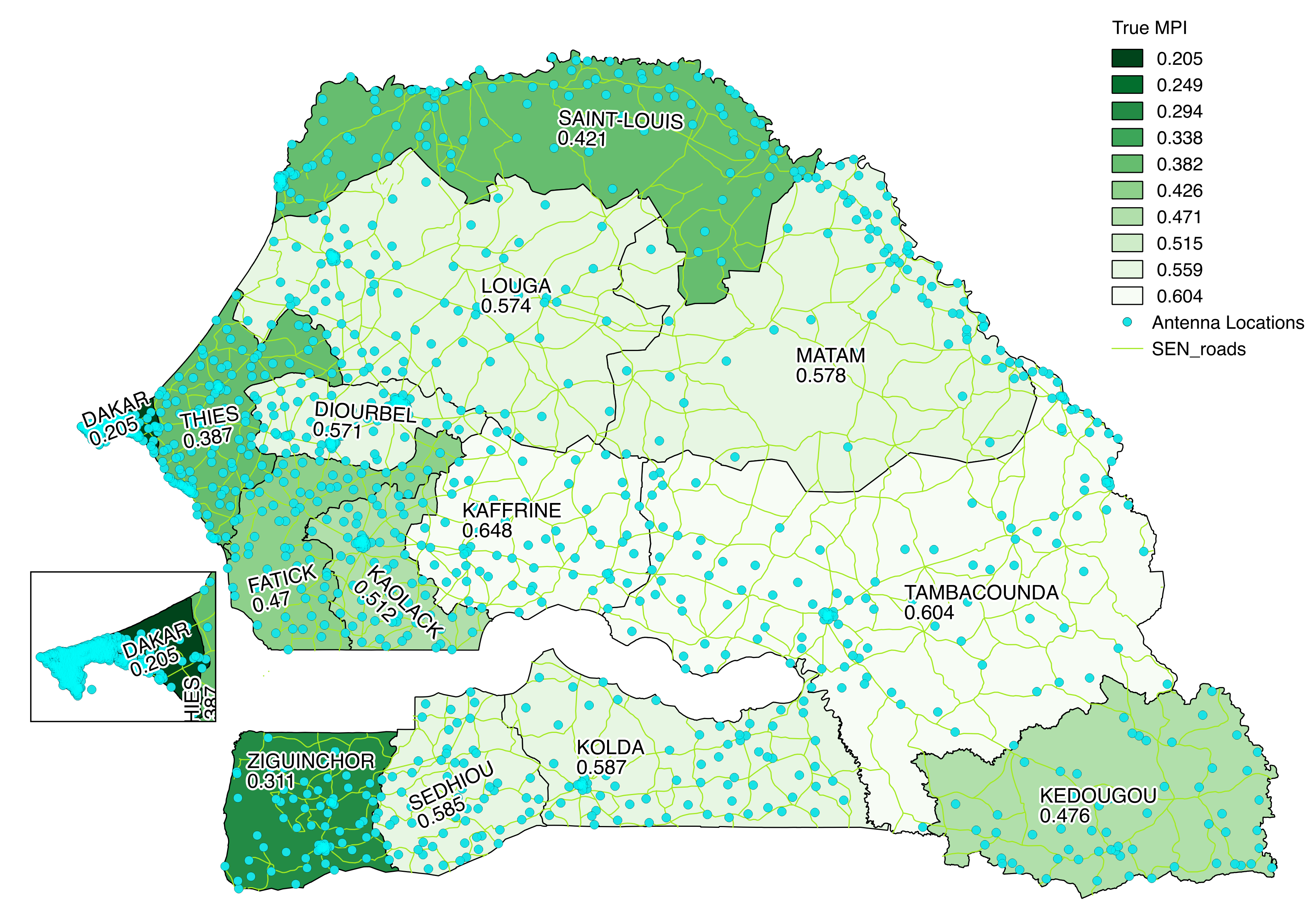}
	\caption{Overlay of Multidimensional Poverty Index (MPI) and cell phone site locations for 14 regions of Senegal.}
	\label{fig:mpimapsites}
\end{figure*}

%Besides the MPI, we also accessed the individual indicators. 
The MPI data is available at region level for each of the 14 regions in Senegal. Figure~\ref{fig:mpimapsites} depicts the latest (2011) poverty map of Senegal.
%
%\begin{figure}[h]
%	\centering
%	\includegraphics[width=0.5\textwidth]{./images/mpimap.pdf}
%	\caption{Multidimensional Poverty Index (MPI) map for 14 regions of Senegal.}
%	\label{fig:mpimap}
%\end{figure}
%

\section{Contributions} 

The main contributions of our work are two-fold. First we construct a virtual network for Senegal from cellular-communication data (Dataset 1), identify network-theoritic measures that correlate well with poverty indicators, learn a model that predicts poverty at a finer resolution, and finally build a poverty map for Senegal at an arrondissement level. Second, we learn a model solely based on the  relationship of aggregated user behavior (Dataset 3) with poverty indicators, and generate a poverty map at a finer resolution. 

Here are detailed technical contributions of our work:

\begin{itemize}
\item We construct a virtual network for Senegal, which is defined as a who-calls-whom network from the mobile communication data. Intuitively, a virtual network quantifies the mobile connectivity, and accessibility to the population. It signifies the macro-level view of connections or social ties between people, dissemination of information or knowledge, or dispersal of services.
\item We study Senegal's virtual network to empirically get the most important spatial regions. We assign each region a unique score based on its importance in the virtual network. We find that a network theoretic based measure, called centrality, provides a strong correlation with the poverty index. The more important the region, the less poor it is on the poverty index.
\item As MPI is a composite index, we find how well each of the component of MPI correlate with the importance of the regions.
%\item Because of the universal nature of mobile phones, we have cellular data at any spatial granularity. Using this data, through our conceptualization, we can construct a virtual network, and get the centrality based scores for any spatial granularity. 
\item We apply linear regression to learn a relationship between centrality and the components of the poverty indicators. Our model is, then, used to estimate the poverty at a finer spatial resolution of arrondissements. We show our predicted region-level poverty map, and validate that it correlates well with the the true poverty map.% We show the arrondissement-level poverty map.%Using provide estimate models that predict the individual components of the index,
\item We provide an in-depth region-level analysis of the correlation and poverty, and explain the bias caused by the Dakar region as it has very high centrality and very low poverty. 
\item We also attempt to understand, and characterize poverty. Since regions may be poor because of various reasons - information poor, but resource rich; or resource poor, but information rich. To study this, we use the behavior indicators of the users provided in Dataset 3. Some of these indicators are: entropy of contacts, percentage of calls from home, radius of gyration, etc. These indicators characterize individual users based on how they call, move, and interact within the cellular network. We studied their correlation with the poverty index and interestingly, found that the correlation was not biased by Dakar. Using one of such indicators, which has a very strong negative correlation with the poverty index, we learn a model for predicting MPI and construct an arrondissement level poverty map for Senegal.
\end{itemize}
%Show that CDRs have been used to answer socio-economic indicators cite...
%When people make mobile-phone calls, the network generates a call data record (CDR) containing such information as the phone numbers of the caller and receiver, the time of the call and the tower that handled it - which gives a rough indication of the device’s location. This information provides researchers with an insight into mobility patterns. Indeed phone companies use these data to decide where to build base stations and thus improve their networks, and city planners use them to identify places to extend public transport.
\section{Related Work}
Call data records (CDR) allow a view of the communication and mobility patterns of people at an unprecedented scale. In the past, several researchers have used CDR data to understand human mobility~\cite{Candia:2008,Calabrese:2010}. However, there has been limited work in understanding relation between CDR data and poverty~\cite{Eagle:2010,Soto:2011,Martinez:2013,Smith:2013}.  Eagle et al.~\cite{Smith:2013} correlated the diversity in communication with socioeconomic deprivation and found a strong positive correlation between the diversity in calling patterns and socioeconomic deprivation in England. The closest work to that presented here is by Smith et al.~\cite{Smith:2013} who have calculated a number of features like introversion, diversity, residuals, and activity to find their correlations between poverty index (MPI) of C\^{o}te d'Ivoire, and further build a finer granularity poverty index based on the feature that gave the best correlations. Our work is different in the following aspects:
\begin{inparaenum}[a).]
	\item we study the virtual communication network using a rigorous network science approach and find that centrality based measures which focus on the importance or influence of nodes provide a better correlation with the poverty index,
	\item we treat MPI as a composite index and provide estimate models that predict the individual components of the index,
	\item we provide an in-depth region wise analysis of the correlation and discover the bias caused by the Dakar region because of its unique nature in Senegal, and
	\item we also compare the finer level poverty maps generated using the network centric approach with a human behavior based method.
\end{inparaenum}
Our work on relating human behavior with MPI is motivated from work done in the past in which behavior indicators are extracted from CDR data and used to predict the socioeconomic indicators of a region~\cite{Soto:2011,Martinez:2013}. Specifically, Soto et al.~\cite{Soto:2011} have proposed a Support Vector Machine model which uses 279 features (calling behavioral, mobility, and social) extracted from an individual users’ CDR to predict the socioeconomic levels at a census region. However, the predictive model requires knowledge of finer granularity poverty data and partial knowledge of a user's home information. Instead, we use the 33 indicators provided in the D4D challenge data and provide a methodology to correlate the indicators at region and arrondissement level without re quiring additional information about the users.

\section{Senegal's Virtual Network}

We define the physical network of a country, as composed of transportation landscape like road, railways, ports, which are the nation's arteries that fuel its economic growth. With the burgeoning growth of mobile communication, we define a virtual network of a country, that describes who-calls-whom network. Calls are placed for a variety of reasons including request of resources, information dissemination, personal. The call data records (CDR), provided by the Orange, provides an interesting way to characterize and understand the virtual network of Senegal. 

While the physical network determines how people move, and goods are transported, virtual network determines how information or knowledge flows. Currently, a good portion of the information and services are dispersed virtually. While the physical network is limited by the inherent capacity of the roads, and railway network, the virtual network is dynamic. Due to people's mobility across spatial regions, and the ubiquitous nature of cellular technology, both physical and virtual networks interact creating complex dynamism. For a holistic understanding of any complex phenomenon, we need to understand both the networks. %Thus, spatial regions are not just topological structures but are dynamical systems with interactions between them. 

Static maps are easy to get, but how to get the virtual network. Existing gravity models can provide an estimate of the flow (with the knowledge of a constant), however such estimates are static and over-reliant on the spatial proximity between sites. The CDR data, however, provides the actual measure of the information flow at a finer spatial and temporal resolution. We construct a virtual network of Senegal from the CDR data. Such a network is generic, and can be used for understanding multiple phenomenon involving dynamic interactions with the physical network, like e-health (while the physical network determines where disease spreads next, virtual network determines how it can be contained by proper dissemination of preventive knowledge), e-education, and e-commerce.

\subsection{How to construct the Virtual Network}
To construct the virtual network, we need two entities: spatial regions, where calls are originated from or are received in; and virtual paths that signify communication among them. %The spatial regions  and the virtual paths are the edges connecting the nodes. 

In virtual network for Senegal, the spatial regions correspond to administrative areas (that can be arrondissement, departments or regions), and the virtual path between each pair of nodes corresponds to the volume of mobile communication (number of calls and texts) between them during the whole year of 2013. 

%\subsubsection{Data and Methods}
For this study, we used the hourly {\em antenna-to-antenna} traffic available for 1666 cell phone towers (sites) to measure communication between sites for 2013 (Dataset 1), as follows:
\begin{itemize}
	\item Create an information flow matrix at site-level $M^s$ with 1666 rows and 1666 columns, such that the entry $M^s_{ij}$ denotes the number of calls and texts exchanged between site $i$ and $j$ during the whole year. Each entry $M_{ij}$ represents the calls and texts originated at site $i$ and received at site $j$.
	\item  To get the arrondissement level information flow, we ``coarsen'' the site to site matrix into a $124 \times 124$ matrix $M^a$, such that the entry $M^a_{ij}$ denotes the total number calls and texts originated at all sites in arrondissement $i$ and received at all sites in arrondissement $j$.
	\item To get the region level information flow, we ``coarsen'' the arrondissement to arrondissement matrix into a $14 \times 14$ matrix $M^r$, such that the entry $M^r_{ij}$ denotes the total number calls and texts originated from all sites in region $i$ and received at all sites in region $j$.
\end{itemize}

\begin{figure*}[t]
	\centering
	\includegraphics[width=0.7\textwidth]{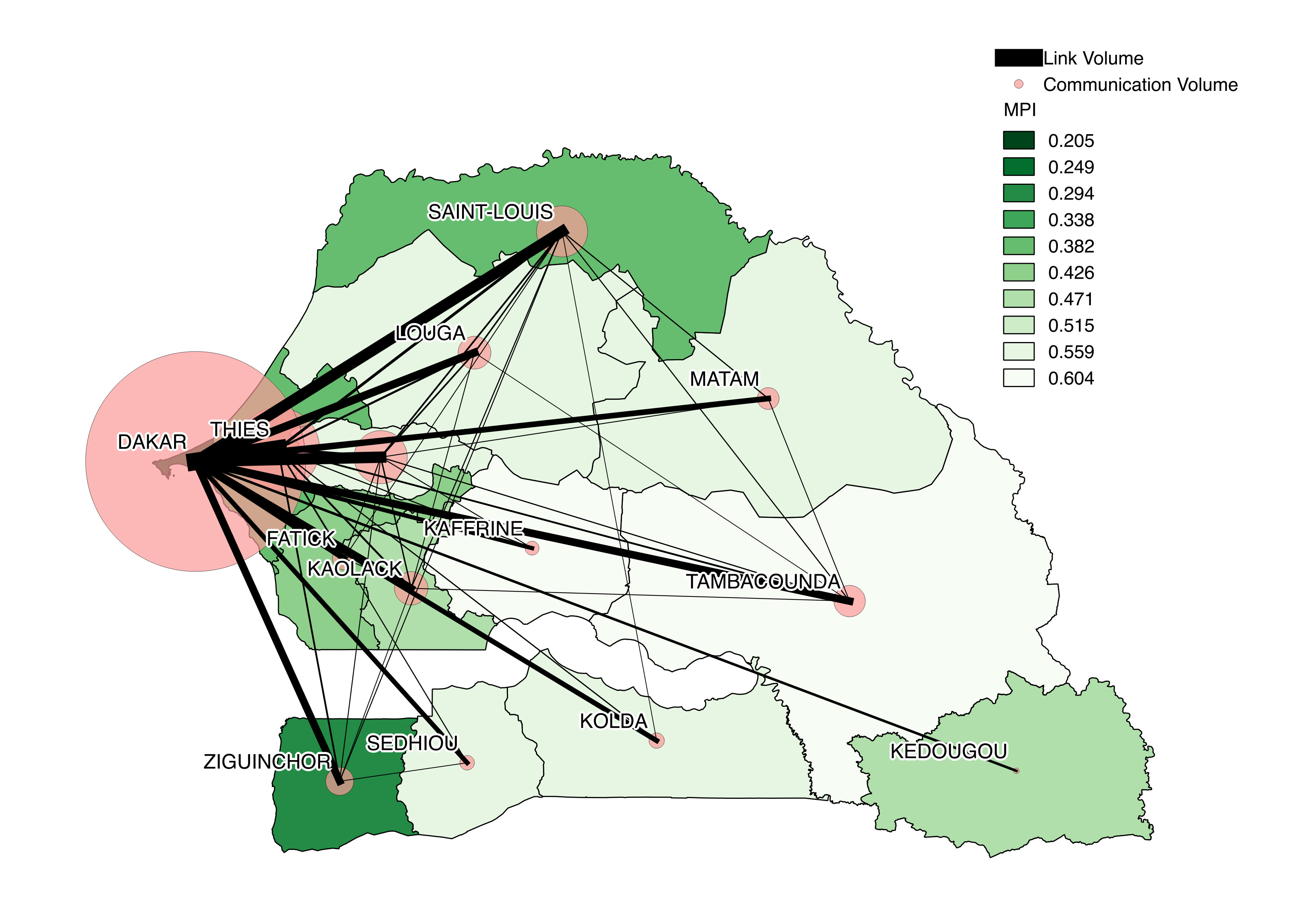}
	\caption{Virtual network for Senegal at Region-level with MPI (Multi-dimensional Poverty) as an overlay. Thickness of links indicate the volume of calls and texts exchanged between a pair of regions. Size of the circle at each region indicates the total number of incoming and outgoing calls and texts for the region. Note that regions with plenty of strong links have lower poverty, while poor regions look isolated.}
	\label{fig:regMap}
\end{figure*}
The resulting virtual network is shown in Figure~\ref{fig:regMap}. On the map, each region is depicted by the latitude and longitude of its geographical centroid. In the graph, the size of each node denotes the total number of incoming and outgoing calls and texts for the region for the entire year. The thickness of the link indicates the volume of calls and texts exchanged between the corresponding pair of regions. Looking at the map, we see that regions, e.g., Dakar, Thies and Ziguinchor, which have low MPI are important nodes in the virtual communication network. On the other hand, regions with high MPI, e.g., Kaffrine and Kolda are not well connected with other regions. However, there are regions that are well-connected but are poor (e.g., Tambacounda) and regions which are poorly connected but are relatively less poor (e.g., Kedougou). This indicates that poverty is a complex phenomenon and needs to be understood from multiple perspectives like relationships with bordering countries, unique geographical settings, etc.
\subsection{Virtual Network and Poverty Analysis}
Figure~\ref{fig:regMap} shows a relationship between the importance of a region in the virtual network with the poverty index. This motivates us to find a quantitative measure of the importance of the region.

As a standard nomenclature, in a network the spatial regions are called nodes, and virtual paths are called edges. Network analysis involves extracting some quantitative measures or properties, associated with the structure of the network, nodes and/or edges.
A popular measure is the relative importance of the nodes in the network. This measure is referred to as centrality~\cite{newmanGraph}. It identifies the most \textit{important} nodes in a network, and assigns a quantitative score to each node. Since importance can have many definitions ranging from nodes being central or cohesive, there are several centrality measures for networks. To calculate the measure we normalize the raw communication matrix ($M^r$) as discussed below.
 
\subsubsection{Normalization of Information Flow Matrix} While previous researchers~\cite{Smith:2013} have used the information flow matrix $M^r$ to characterize the virtual network, we first normalize the matrix to account for the disparity in the population across regions, especially the strong influence of Dakar owing to its relatively high population share (~22.9\%). We construct a normalized information matrix $\hat{M}^r$ as follows:
\begin{equation}
\widehat{M}^r_{ij} = \frac{M^r_{ij}*d_{ij}}{n_in_j}
\label{eqn:1}
\end{equation}
where $n_i$ (and $n_j$) is the number of cell-phone towers in region $i$ (and $j$) and $d_{ij}$ is the as-the-crow-flies distance between the centroids of regions $i$ and $j$. We normalize the matrix $M^a$ in a similar way.

We use the number of sites ($n_i$) as an indicator of the size of the {\em calling population} in the given region. 
%In fact, $n_i$ is strongly correlated (Pearson correlation - 0.91) with the total population of the region. 
Thus, the value $\frac{n_in_j}{d_{ij}}$ in \eqref{eqn:1} is proportional to the expected number of calls between two regions, which is very similar to the well-known {\em Gravity model}, which has been used in the past to predict the intensity of mobile phone calls between cities~\cite{Krings:2009} ($numcalls \propto \frac{p_ip_j}{d^\alpha_{ij}}$). However, instead of using the population of the two regions ($p_i$ and $p_j$), we use the number of sites to get a estimate of the population \textit{with} mobile phones. We set the exponent $\alpha$ for distance as 1 as it gave the best correlation with poverty level. The normalization procedure removes the impact of regional population and spatial distance on the information flow and measures the {\em residual flow}, assuming that all region have equal population and are equidistant from each other.
%\subsection{Quantitative Analysis of the Virtual Network and Poverty}
\subsubsection{Measures of Importance and their Correlation with Poverty}
\begin{table*}[t]
	{\small
		\begin{center}
			\begin{tabular}{|p{1.5in}|c|c|c|c|c|c|}
				\hline
				\multirow{2}{*}{Measure}		  & \multicolumn{2}{|c|}{$H$} & \multicolumn{2}{|c|}{$A$} & \multicolumn{2}{|c|}{$MPI$}\\
				\cline{2-7}
				& $corr$ & $p$-value & $corr$ & $p$-value & $corr$ & $p$-value\\
				\hline
				PageRank & -0.87 &{\footnotesize $6 \times 10^{-5}$} & -0.81 & {\footnotesize 0.0004} & -0.82 & {\footnotesize 0.0003}\\
				\hline
				Eigenvalue Centrality & -0.83&{\footnotesize 0.0002}& -0.80 & {\footnotesize 0.0005}& -0.79& {\footnotesize 0.0007}\\
				\hline
				Gravity Residual  & -0.81&{\footnotesize 0.0003} & -0.76 & {\footnotesize 0.0015}& -0.79& {\footnotesize 0.0007}\\
				\hline			
				Introversion & 0.82&{\footnotesize 0.0002}& 0.70 & {\footnotesize 0.0040} & 0.79& {\footnotesize 0.0006}\\
				\hline
				Activity ({\em Normalized})    & -0.81&{\footnotesize 0.0008} & -0.76 & {\footnotesize 0.0003} & -0.79& {\footnotesize 0.0015}\\
				\hline
				Activity ({\em Raw})& -0.80&{\footnotesize 0.0006} & -0.68 & {\footnotesize 0.0075}  & -0.71& {\footnotesize 0.0040} \\
				\hline
			\end{tabular}
		\end{center}
	}
	\caption{\small Pearson's r Correlation of region-wise poverty indicators with communication graph features. $H$ -- Incidence of Poverty, $A$ -- Average Intensity Across the Poor, $MPI$ -- Multidimensional Poverty Index.}
	\label{tab:corr}
\end{table*}
Besides graph-theoretic measures, we also investigated direct features that can be calculated from the raw communication matrix ($M^r$) or the normalized communication matrix ($\widehat{M}^r$) discussed as follows. %Some of these have been used in the past in a similar context~\cite{Smith:2013,Eagle:2010}.
\begin{itemize}
	\item \textbf{Activity}: This feature is a simple aggregate of {\em outgoing flows} from a region ($=\sum_{i \neq j} M^r_{ij}$ for region $i$). We also used a similar feature derived from the normalized matrix ($=\sum_{i \neq j} \widehat{M}^r_{ij}$ for region $i$). Additionally, we investigated other variants such as the count of {\em incoming flows}, {\em within flows}, and {\em total flows} and found similar relationships with the poverty indicators.
	\item \textbf{Eigen Vector and Page Rank Centrality}: This is derived from the normalized matrix $\widehat{M}^r$,and is a measure of the influence of a node in a graph. Eigen vector centrality of a node, $v_i$, is a weighted sum of centralities of all of its outgoing connections:
	\begin{equation}
	x_i = \frac{1}{\lambda}\sum_j \widehat{M}^r_{ij} x_j
	\label{eqn:eigen}
	\end{equation}
	where $\lambda$ is some constant. In matrix notation, this can be written as $\lambda {\bf x} = \widehat{M}^r{\bf x}$, such that ${\bf x}$ is the eigenvector of the matrix $\widehat{M}^r$ corresponding to the leading eigenvalue.
	%
	%It is based on the ideas that not all connections are equal. Thus, %It assigns relative score to a node by giving more weight to its connections to high-influencing nodes, and less weight to its connections to low-influencing nodes. It is b
	%a node that is connected to other important nodes gets higher score, than a node that is connected to nodes of low importance.
	
	\textbf{Page Rank} is a variant of eigen vector centrality and is widely used for ranking websites by search engines such as Google. However, the actual role of Page Rank is to rank nodes in a network based on their importance. It has been noted that the classic eigen vector centrality (See~\eqref{eqn:eigen}) performs poorly for directed networks while the Page Rank measure can handle directed networks better.
	\item \textbf{Gravity Residual}: As shown in \eqref{eqn:1}, each entry of the normalized matrix $\widehat{M}^r$ measures the ``residual'' from node $i$ to $j$ after normalizing for population and spatial distance. We compute the total outgoing residual flow from each node as:
	\begin{equation}
	Residual_i = \sum_j \widehat{M}^r_{ij}
	\label{eqn:residual}
	\end{equation}
	In the past~\cite{Smith:2013}, similar measures have been shown to correlate negatively with MPI, indicating that regions that communicate more are less poor.
	\item \textbf{Introversion}: This measures the tendency of the population within a region to communicate within the region instead of outside. The introversion measure can be calculated as:
	\begin{equation}
	Introversion_i = \frac{M^r_{ii}}{\sum_j M^r_{ij}}
	\label{eqn:introversion}
	\end{equation}
\end{itemize} 
All the above measures give a score for each region in Senegal, based on its relative importance. Further, we study how these measures correlate with the poverty index of the regions. The MPI reflects both the incidence or headcount ratio ($H$) of poverty, i.e., the proportion of the population that is multidimensionally poor – and the average intensity ($A$) of their poverty, i.e., the average proportion of indicators in which poor people are deprived. The MPI is calculated by multiplying the incidence of poverty by the average intensity across the poor ($H \times A$). Hence, we study the correlation of the network features with $H$, $A$, and $MPI$. Table~\ref{tab:corr} shows the {\em Pearson's r correlation} and the corresponding $p$-values. We observe that the various metrics have a strong negative correlation with the $H$ value, which is the headcount ratio of poverty. Similarly, the metrics have a marked negative correlation with $A$, which is  the incidence of poor, and also with MPI of the regions.
%This result emphasis on how the MPI is calculated at a disaggregated level of arr. Naively, assigning all the arr within a region same MPI, does not prove good. Thus, there is a need to generate poverty maps at a finer disaggregated level.

%\subsubsection{Region Wise Analysis of Poverty and Communication Features}
\subsubsection{Strong influence of Dakar region}
\begin{figure*}[p]
	
	\begin{subfigure}{0.45\textwidth}		
		\centering		
		\includegraphics[width=\textwidth]{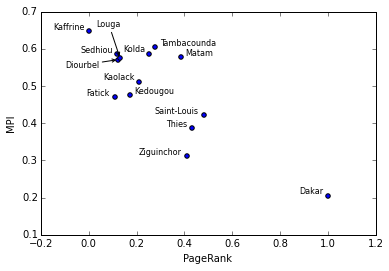}		
		\caption{With Dakar}
	\end{subfigure}
	\begin{subfigure}{0.45\textwidth}
		\centering
		\includegraphics[width=\textwidth]{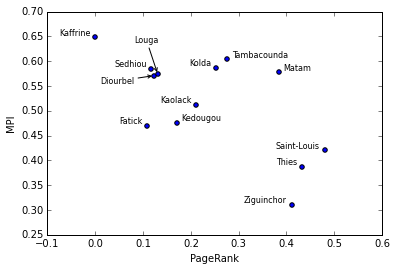}
		\caption{Without Dakar}
	\end{subfigure}
	\caption{Illustrating influence of Dakar on the relationship between MPI and Page Rank for regions.}
	\label{fig:dakarinfluence}
\end{figure*}

Although pagerank exhibits strong correlation with the indicators $H$, $A$, and $MPI$, when we plot the correlation (see Figure~\ref{fig:dakarinfluence}) we see that Dakar has a very unique characteristic of very high centrality, and very low MPI, and occupies a corner in the scatter plot, whereas all other regions are spread at the other corner, with mid-to-high MPI and low-to-mid pagerank. We, then, remove Dakar, to see its effect on the correlation of pagerank with the poverty indicators.  Surprisingly, we lose the high correlation between pagerank and poverty indicators when Dakar is removed.  This is evident from Table~\ref{tab:dakar}. The correlation drops significantly with high p-values. 

%We call it a bias, as Dakar is the only region in Senegal with high centrality, whereas other regions of Senegal have low centrality.
We attribute this to its geopolitical heritage and past history as a port during the colonial times. It is the largest city with 2.47 million people, followed closely by Grand Dakar at 2.35 million. There has been excessive economic activity in Dakar, which makes up more than half of the Senegalese economy in less than 1 percent of the national territory. But sustained economic development, there needs to be de-centralized development focusing on marginalized areas. This fact is also validated by the International Monetary Fund's 2013 report on Senegal~\cite{imf}.  

\begin{table*}[p]
	{\small
	\begin{center}
		\begin{tabular}{|p{1.5in}|c|c|c|c|c|c|}
			\hline
			\multirow{2}{*}{Measure}		  & \multicolumn{2}{|c|}{$H$} & \multicolumn{2}{|c|}{$A$} & \multicolumn{2}{|c|}{$MPI$}\\
			\cline{2-7}
			& $corr$ & $p$-value & $corr$ & $p$-value & $corr$ & $p$-value\\
			\hline
			PageRank with Dakar& -0.87 &{\footnotesize $6 \times 10^{-5}$} & -0.81 & {\footnotesize 0.0004} & -0.82 & {\footnotesize 0.0003}\\
			\hline
			PageRank without Dakar& -0.68&{\footnotesize 0.01}& -0.65 & {\footnotesize 0.016}& -0.64& {\footnotesize 0.018}\\
			\hline
		\end{tabular}
	\end{center}
	}
	\caption{\small Pearson's r Correlation of $H$, $A$ and $MPI$ with Pagerank of the regions considering Dakar and NOT considering Dakar.}
	\label{tab:dakar}
\end{table*}

\subsubsection{Generating Finer Resolution Poverty Maps}

To illustrate how to derive finer resolution poverty maps (at department or arrondissement levels), we use pagerank from Table~\ref{tab:corr} to predict $H$ and $A$. We learn two linear models using ordinary least squares regression to predict the {\em Incidence of Poverty} ($H$) and {\em Average Intensity across Poor} ($A$). The learnt models are:

\begin{eqnarray}
\tilde{H}_i & = & -708.32 \times PageRank_i  + 131.94\label{eqn:lm1}\\
\tilde{A}_i & = & -346.66 \times PageRank_i + 84.58\label{eqn:lm2}
\end{eqnarray}

Finally, we combine the two estimates to predict the MPI as:

\begin{equation}
\widetilde{MPI}_i = \frac{\tilde{H}_i}{100} \times \frac{\tilde{A}_i}{100} \label{eqn:predmpi}
\end{equation}

\begin{figure}[h]
	\centering
	\includegraphics[width=0.5\textwidth]{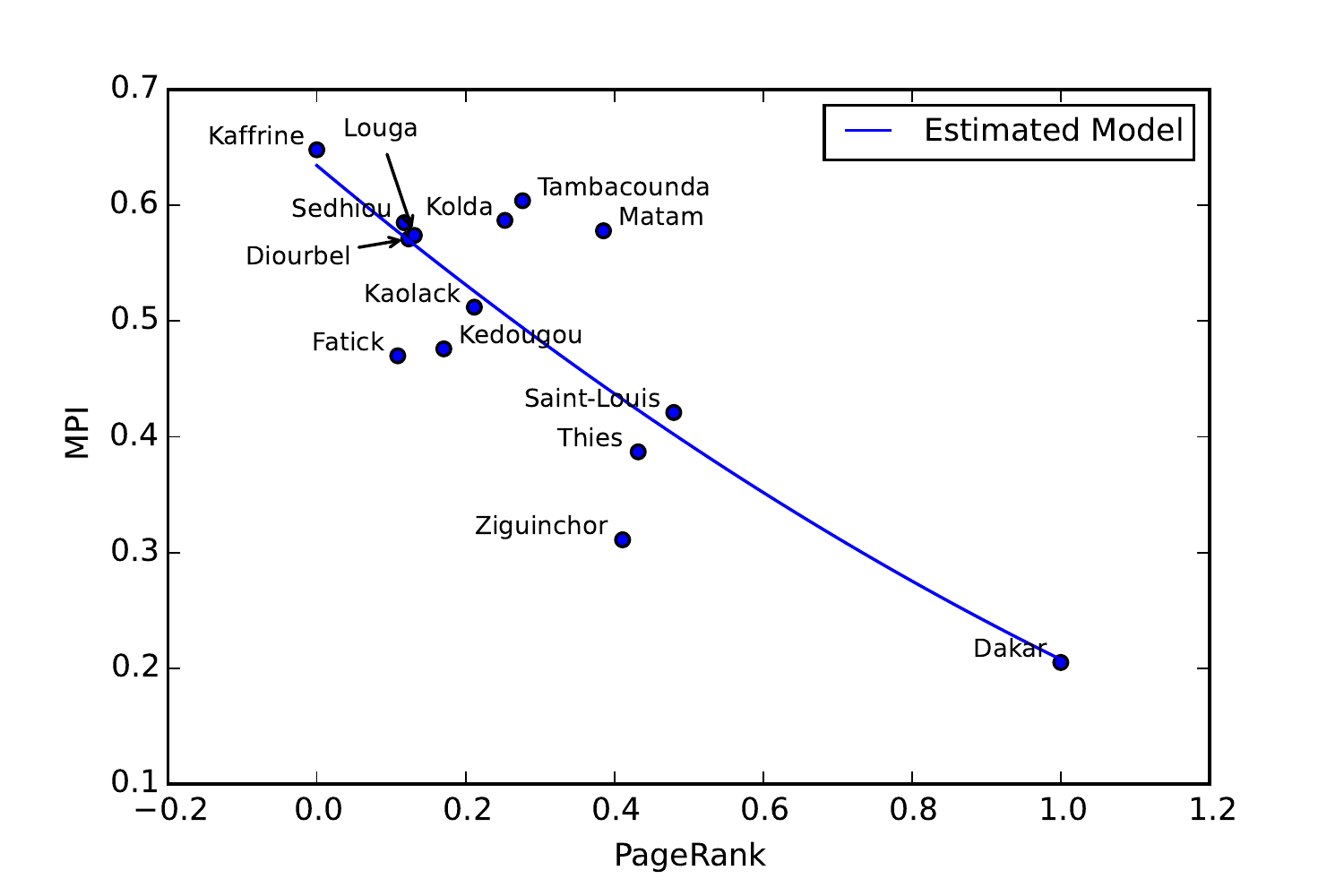}
	\caption{Estimated model for predicting MPI using the Page Rank feature.}	
	\label{fig:scatterwithestimate}	
\end{figure}

The estimated model for predicted MPI is shown in Figure~\ref{fig:scatterwithestimate}. Using this model, we can estimate the MPI at a finer spatial resolution, as long as we can compute the pagerank for the target spatial areas.

\begin{figure*}[p]
	\centering
	\includegraphics[width=0.7\textwidth]{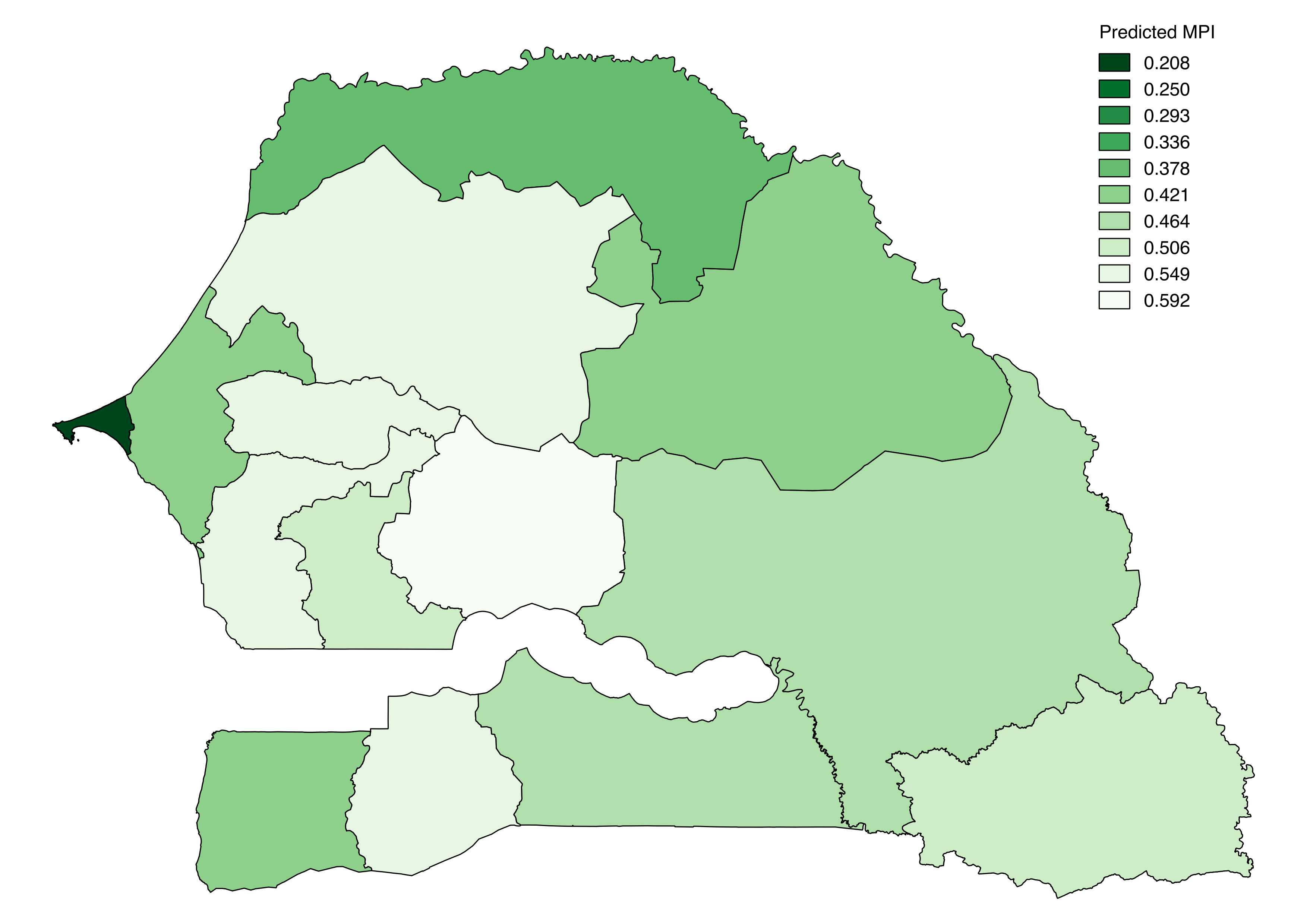}
	\caption{Predicted region level Poverty map using the virtual network.}
	\label{fig:predictedregmpi}
\end{figure*}

The region level predicted MPI map is shown in Figure~\ref{fig:predictedregmpi}. Note its similarity with the true MPI Map of Senegal in Figure~\ref{fig:mpimapsites}.
\begin{figure*}[p]
	\begin{subfigure}{0.5\textwidth}		
		\centering		
		\includegraphics[width=\textwidth]{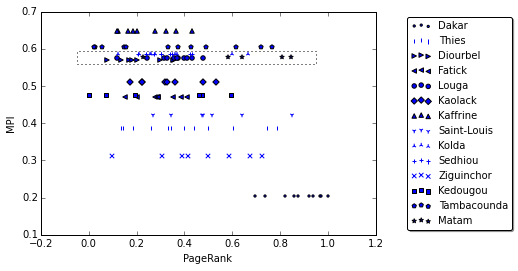}		
	\end{subfigure}
	\begin{subfigure}{0.5\textwidth}
		\centering
		\includegraphics[width=\textwidth]{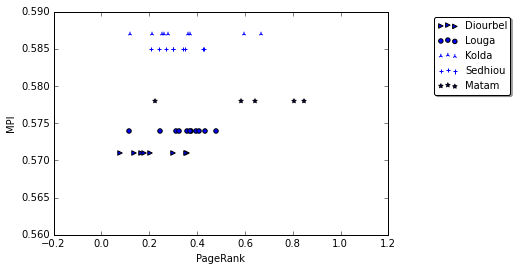}
	\end{subfigure}
	\caption{Visual depiction of what happens when MPI is calculated at a region level. All arrondissements within each region are assigned the same poverty, but they have varying centrality measures, signifying importance! Thus, need to generate finer povery maps for targeted eradication of poverty. The dashed box in the top panel is expanded in the bottom panel.}
	\label{fig:scatter_arr_pagerank}
\end{figure*}
\begin{figure*}[p]
	\centering
	\includegraphics[width=0.7\textwidth]{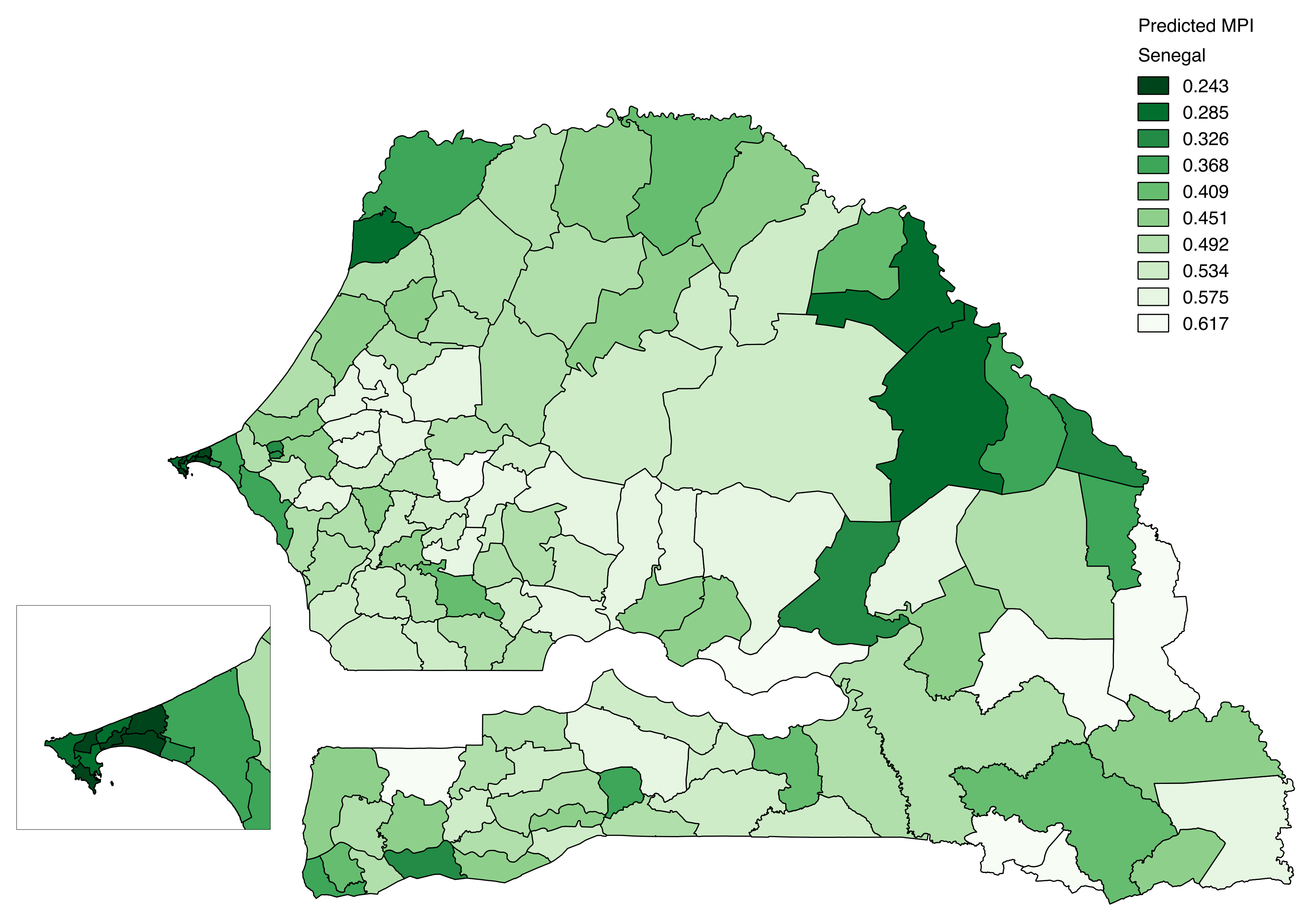}
	\caption{Predicted Arrondissement level Poverty map using the virtual network.}
	\label{fig:predictedmpi}
\end{figure*}

Figure ~\ref{fig:scatter_arr_pagerank} motives the need for finer granularity poverty maps than regions. We can observe a significant variability in the centrality measure across arrondissements within the same region. This indicates that a region has varying levels of poverty. For targeted distribution of economic resources, we need finer level poverty maps than regions.

To generate the arrondissement level poverty map, we first compute its Page Rank using the normalized matrix $\widehat{M}^a$.

Then we use the models in~\eqref{eqn:lm1}--\eqref{eqn:predmpi} to predict the MPI for each arrondissement. The predicted poverty map is shown in Figure~\ref{fig:predictedmpi}. It is interesting to see that regions are composed of arrondissements with varying poverty index. 
\section{Correlating Behavioral Indicators of users and Poverty}\label{sec:behavior}
%In previous section we used CDR data to understand the virtual communication network of Senegal and its relationship with poverty. 
In this section, we study how user behavioral statistics, gathered from cellular-communications, correlate with the poverty indicators. Can this relationship be learnt as a model to generate poverty maps at a finer resolution?
%we use the CDR data to understand the individual behavior and how it maybe used as an indicator of the poverty level of the individual or the community to which he/she belongs. Of particular interest is the question of how the resulting poverty map from human behavior analysis compares to the map obtained using the virtual network analysis.

As previous researchers have shown~\cite{Soto:2011,Martinez:2013}, human behavioral information extracted from CDR data can be used to measure the socio-economic development of a region. %In particular, human mobility data, extracted from call records, has been shown to be strongly linked to the socio-economic status of the target population~\cite{Soto:2011}. But besides mobility, CDR data also allows characterizing social and consumption behavior of individuals, which have been shown to be related to the socio-economic levels.
We study the relationship of several human behavior indicators extracted from CDR data with MPI with the goal of identifying key indicators which can then be used to predict MPI at a finer spatial resolution.
\subsection{Data}
For this study we use the one year of coarse-grained mobility data available at arrondissement level for 146,352 users (referred to as Set3 data). For each user, the data records the location (at arrondissement level) and time (at hourly level) at which the user makes a call or sends a text. Additionally, the data also contains a monthly set of 33 behavioral indicators which capture calling/texting patterns (14), mobility patterns (6), and social behavior (13) of each user.
\subsection{Aggregating User Behavior}
For each user, we compute the median of the 12 monthly values, for each of the 33 indicators. To relate these individual level indicators with region level MPI data, we need to assign a ``home region'' to each user. This information is not provided in the data set. We employ the following localization procedure to assign an arrondissement (and a region) to each user in the sample of 146,352.
\subsubsection{Localization of Users}
For each user we consider the calls made between 8 PM and 12 PM on each of the 365 days of the year. We measure the following quantities:
\begin{enumerate}
	\item $d_i$: Fraction of days (out of 365) the user $i$ made at least one call between 8 PM and 12 PM in the whole year.
	\item $a_i$: The integer id (between 1 and 123) of the arrondissement that the user $i$ called most frequently from during those hours.
	\item $c_i$: Fraction of total calls made by the user $i$ between 8 PM and 12 PM from the arrondissement $a_i$.
\end{enumerate}
The arrondissement $a_i$ is assigned as the ``home arrondissement'' for user $i$ and the corresponding region is the ``home region''. We filter out individuals with {\em insufficient} (low $d_i$) or {\em ambiguous} (low $a_i$) information by ignoring all users for whom $d_i \le 0.5$ and $c_i \le 0.95$, i.e., we only considered those users who made a call in the night at least half of the days in the year and who called from a single arrondissement 95\% of the time. After filtering the sample contained 33,323 individuals (23\% of the original sample).

\begin{figure}[h]
	\centering
	\includegraphics[width=0.5\textwidth]{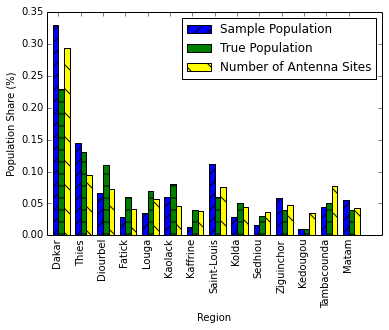}
	\caption{Comparing the region-level population share for the indicator sample and the true population.}
	\label{fig:part3popshare}
\end{figure}

To verify that the filtered sample represents the entire country we compare the region wise distribution of the individuals with true 2011 population share and the region wise share of the number of cell phone antenna sites in Figure~\ref{fig:part3popshare}. We observe that while our filtered set of users oversamples from some of better developed regions of Senegal, the distribution approximates the population as well as number of sites (which in turn is an indicator of the number of mobile users) across regions.
\subsubsection{Region-Level Behavioral Indicators}
To compute the indicators for each region, we consider all users assigned to that region. For each indicator, we compute the median value for each indicator. Thus we obtain 33 median indicators for each region.
\subsection{Aggregated Behavioral Indicators and Poverty Analysis}
For each indicator we compute the Pearson's r correlation between the region level median value for that indicator and MPI. Out of 33 indicators, 11 had an absolute correlation of 0.90 or greater with $p$-value $< 0.00001$. We chose one of these indicator variables with the strongest correlation with MPI -- {\em Percentage Initiated Conversation} (PIC). PIC had a negative correlation of -0.93 ($p$-value = $2\times 10 ^{-06}$) with MPI. Additionally, this indicator (as well as all other indicators) were not significantly influenced by Dakar (correlation without Dakar = -0.89, $p$-value = $4\times 10^{-05}$).

\begin{figure*}[t]
	\centering
	\includegraphics[width=0.75\textwidth]{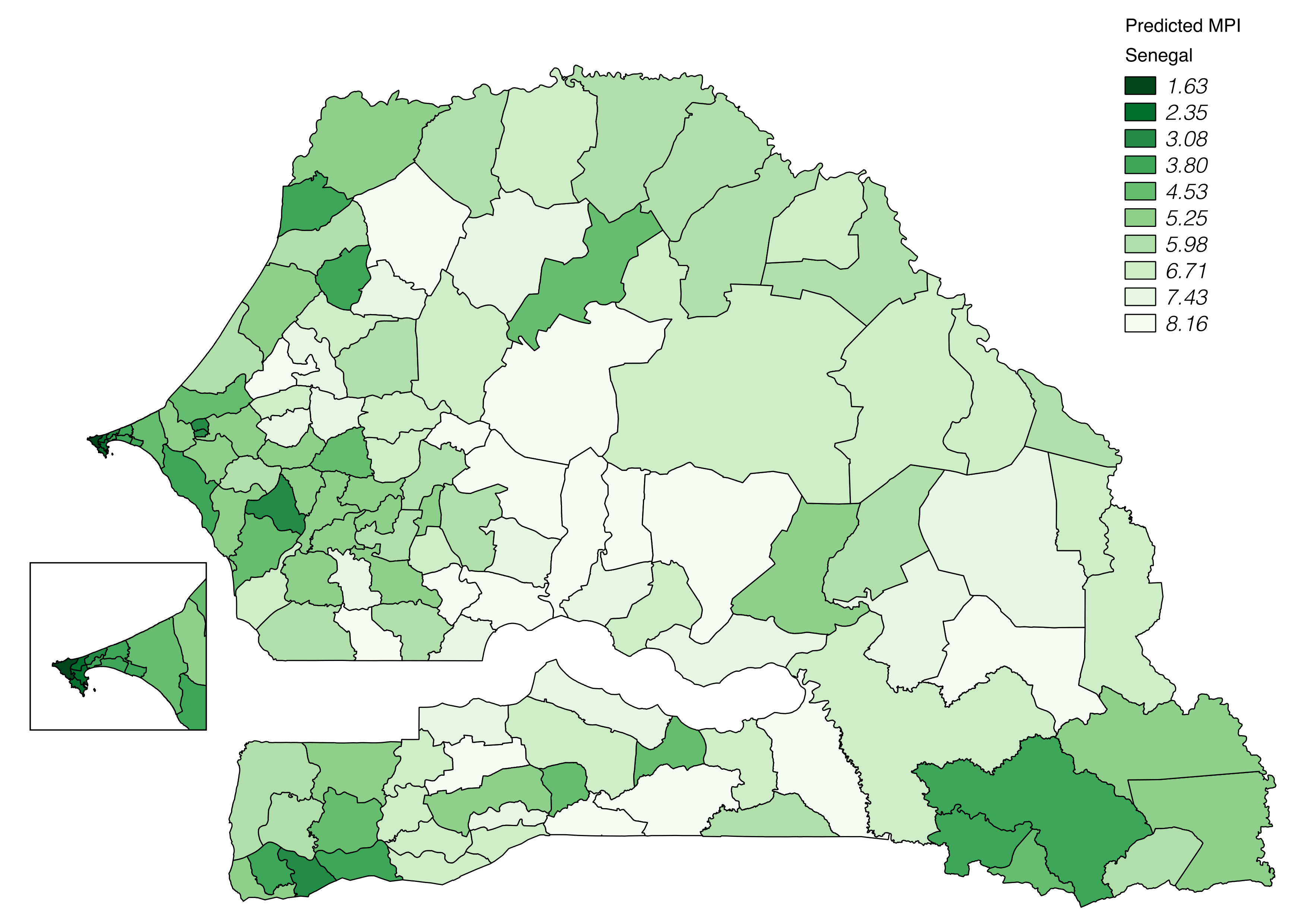}
	\caption{Predicted arrondissement-level map of MPI for Senegal using Behavioral Indicators.}
	\label{fig:predictedmpibehavior}
\end{figure*}

Result for PIC indicate that in regions with low MPI users tend to initiate more call/texts than the users belonging to regions with higher MPI. Similar to previous approach, we found the linear regression models to predict incidence of poverty ($H$), average intensity across poor ($A$) and eventually, the MPI for a given geographical region. The parameters of the linear models are:
\begin{eqnarray}
\tilde{H}_i & = & -302.65 \times PIC_i  + 119.35\label{eqn:lm3}\\
\tilde{A}_i & = & -151.53 \times PIC_i + 78.84\label{eqn:lm4}
\end{eqnarray}
\begin{figure}[h]
	\centering
	\includegraphics[width=0.5\textwidth]{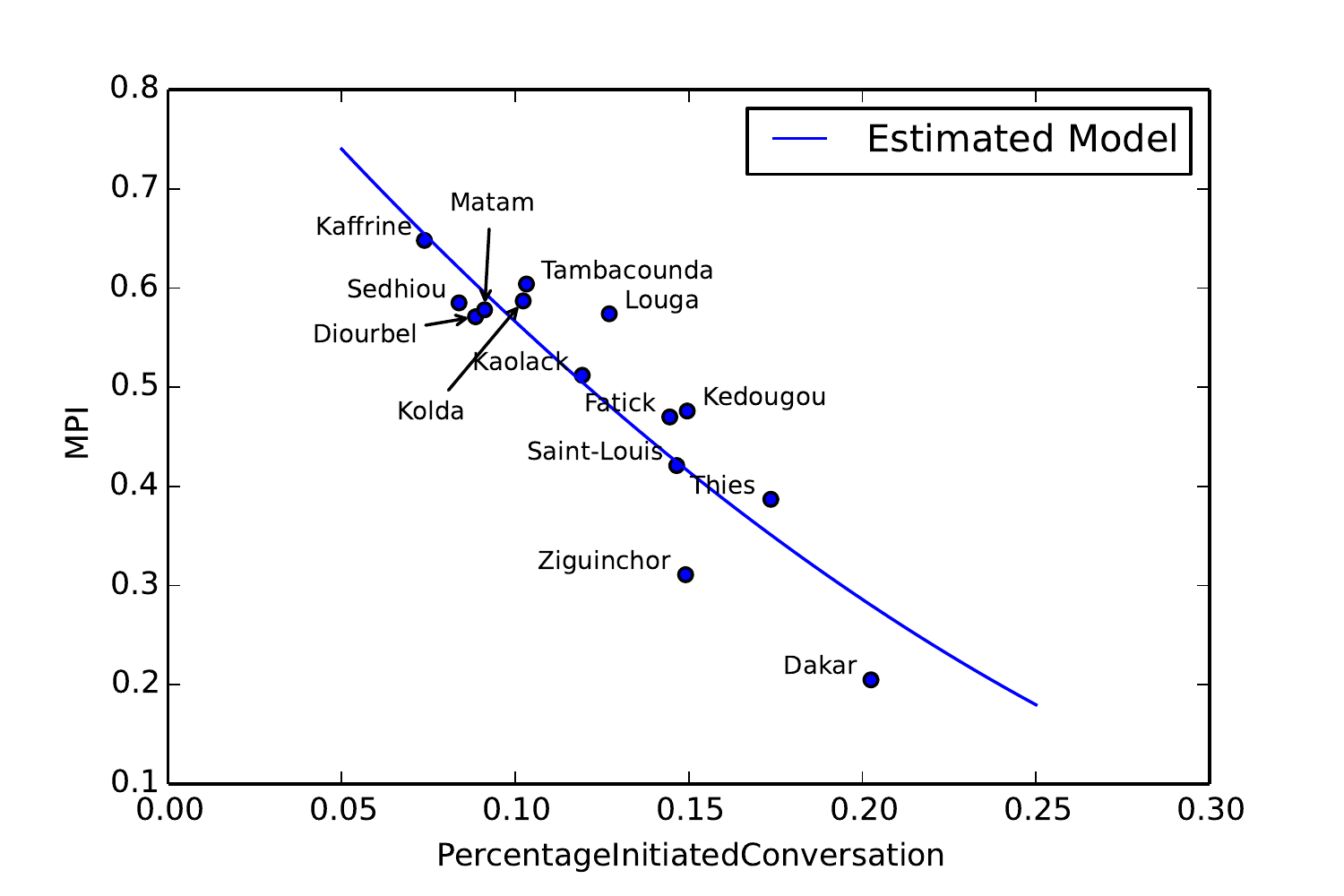}
	\caption{Estimated model for predicting MPI using the Page Rank feature.}	
	\label{fig:scatterwithestimatebehavior}	
\end{figure}
The MPI is calculated by multiplying the two estimates (See~\eqref{eqn:predmpi}). The estimated model for MPI using~\eqref{eqn:lm3} and~\eqref{eqn:lm4} is shown in Figure~\ref{fig:scatterwithestimatebehavior}. Using a similar procedure as discussed in previous section, we generate an arrondissement level poverty map for Senegal as shown in Figure~\ref{fig:predictedmpibehavior}.

\section{Conclusions}
We analyze the virtual network for Senegal, constructed from call data records (CDR) in the context of understanding poverty. We propose a novel methodology to construct such networks at varying spatial resolutions, such as regions or arrondissements. We apply network centric methods, such as centrality, to measure the importance of each node in the virtual network, where the node either corresponds to one of the 14 regions or 123 arrondissements in Senegal. We show strong correlation of centrality and other measures with the poverty index of the region level nodes.

Since Multi-dimensional Poverty Index (MPI) as a composite of two individual indices, we learn a model that correlates poverty with each of the indicators. This allows us to learn a better relationship between the network centric measures and MPI. We provide an in-depth region-level analysis of the correlation between centrality and MPI and discover a bias induced by the Dakar region and further analyze the cause of such bias. We provide an approach to utilize the user behavioral indicator data to understand their relationship with the MPI. This is the first time such analysis has been done to understand MPI. Through our analysis we discover indicators which are not only strongly correlated with MPI at region level (0.92 Pearson's r correlation) but also are not biased by any particular region, as was observed for the centrality measures.

Since poverty is a complex phenomenon, poverty maps showcasing multiple perspectives, such as ours, provide policymakers with better insights for effective responses for poverty eradication. Poverty maps at arrondissement and department levels, or at any spatial levels, will enable targeted policies for inclusive growth of all the regions in Senegal. The poverty maps generated using the behavioral indicators can be used to focus policies for certain demographics of the society that are specially vulnerable to poverty, such as women and specific ethnic groups.

\section*{Acknowledgment}
We gratefully thank Professor H.~C.~Pokhriyal, at University of Delhi, India and Dr.~Varun Chandola, SUNY Buffalo for many thoughtful discussions. 
\bibliographystyle{ieee}
\bibliography{refs}

\end{document}